\documentclass[aps,prf,reprint,showpacs,superscriptaddress]{revtex4-1}

\usepackage{graphicx}
\usepackage{subfigure}
\usepackage{color}
\usepackage{amsmath}
\usepackage{lipsum}
\usepackage{amssymb}
\usepackage{epstopdf}
\usepackage[linktocpage,colorlinks=true,linkcolor=blue,citecolor=blue,breaklinks=true]{hyperref}
\usepackage{verbatim}
\usepackage{breakcites}
\usepackage{mathrsfs}
\usepackage[version=3]{mhchem}

\newcommand{\be}{\begin{equation}}
\newcommand{\ee}{\end{equation}}

\begin{document}

\preprint{}

\title{Penetrative Convection at High Rayleigh Numbers}

\author{Srikanth Toppaladoddi}
\affiliation{Yale University, New Haven, USA}
\affiliation{All Souls College, University of Oxford, Oxford, UK}
\affiliation{Department of Physics, University of Oxford, Oxford, UK}
\affiliation{Mathematical Institute, University of Oxford, Oxford, UK}

\author{John S. Wettlaufer}
\affiliation{Yale University, New Haven, USA}
\affiliation{Mathematical Institute, University of Oxford, Oxford, UK}
\affiliation{Nordita, Royal Institute of Technology and Stockholm University, Stockholm, Sweden}
\email[]{john.wettlaufer@yale.edu}

\date{\today}

\begin{abstract}
We study penetrative convection of a fluid confined between two horizontal plates, the temperatures of which are such that a temperature of maximum density lies between them. The range of Rayleigh numbers studied is $Ra = \left[10^6, 10^8 \right]$ and the Prandtl numbers are $Pr = 1$ and $11.6$. An evolution equation for the growth of the convecting region is obtained through an integral energy balance. We identify a new non-dimensional parameter, $\Lambda$, which is the ratio of temperature difference between the stable and unstable regions of the flow; larger values of $\Lambda$ denote increased stability of the upper stable layer.  We study the effects of $\Lambda$ on the flow field using well-resolved lattice Boltzmann simulations, 
and show that the characteristics of the flow depend sensitively upon it. For the range $\Lambda = \left[0.01, 4\right]$, we find that for a fixed $Ra$ the Nusselt number, $Nu$, increases with decreasing $\Lambda$. We also investigate the effects of $\Lambda$ on the vertical variation of convective heat flux and the Brunt-V\"{a}is\"{a}l\"{a} frequency. Our results clearly indicate that in the limit $\Lambda \rightarrow 0$ the problem reduces to that of the classical Rayleigh-B\'enard convection.

\end{abstract}

\maketitle

\section{Introduction}

Penetrative convection refers to situations where a gravitationally unstable layer of fluid advances into a stable layer of fluid \citep{veronis1963, spiegel1972}. The motion of the fluid in the unstable layer is typically driven by a source of heat. Penetrative convection is relevant in both astrophysical and geophysical settings \cite[e.g.,][]{veronis1963, saunders1962, zahn1977}, with typical examples of the former being the interaction between convective and radiative zones in stars \cite{brummell2002, Shravan:2016} and of the latter being the destruction of the near-ground stable layer in the atmosphere due to radiative heating from the ground \cite{deardorff1969, tennekes1973} and the deepening of the upper ocean mixed layer due to surface cooling or formation of sea ice \cite{farmer1975, turner1979}.

For concreteness we study penetrative convection in water, which has a density maximum at $T_M = 4$ $^{\circ}$C. If the upper surface of a column of water is maintained at a temperature below $T_M$ and the lower surface is maintained at a temperature above $T_M$, the layer of fluid with temperature below $T_M$ is stably stratified and the layer with temperature above $T_M$ is unstably stratified. 
As the value of $Ra$ for the unstable layer increases, convection will be initiated, which then leads to the entrainment of the fluid from the stable layer and hence growth of the convecting region. 

The first stability analysis of penetrative convection was carried out by Veronis \cite{veronis1963}, who considered a column of water the bottom of which is maintained at $0$ $^{\circ}$C and the top of which is maintained at a temperature greater than $4$ $^{\circ}$C, along with stress-free conditions for velocity. From a linear stability analysis of the Boussinesq equations he found that as the temperature of the upper boundary increases, the critical Rayleigh number ($Ra_c$) for the unstable layer decreases from its value for the classical Rayleigh-B\'enard problem, reaching a minimum before attaining an asymptotic value. Veronis \cite{veronis1963} argued that this  behavior of $Ra_c$ is due to three competing factors: (1) The presence of a stable layer relaxes the upper boundary condition, thereby allowing the flow in the unstable region to reach an ``optimum" state. As the thickness of the stable layer increases with the top-plate temperature, higher values of the temperature are preferred; (2) The number of cells in the vertical increases with increasing temperature, with the cell in the stable layer deriving its energy from the flow in the unstable layer. Hence, to minimize this energy loss, lower values of the top-plate temperature are preferred; (3) The available potential energy increases up to a top-plate temperature of $8$ $^{\circ}$C, and does not change with any further increase in the temperature, thereby clearly favoring a top-plate temperature of $8$ $^{\circ}$C. A combination of these three factors results in $Ra_c$ attaining a minimum at $6.7$ $^{\circ}$C. Veronis \cite{veronis1963} also discovered that convection could set in at subcritical values of Rayleigh number, because any finite-amplitude disturbance that mixes water layers above and below the level of maximum density leads to the creation of a deeper unstable layer, thereby favoring onset of convection.

The first experimental study of penetrative convection was by Townsend \cite{townsend1964}, who examined turbulent natural convection over a layer of ice. The bottom surface of the tank was ice covered and the upper free surface was maintained at a temperature of $25$ $^{\circ}$C. We estimate that the $Ra$ in his experiments, based on the total depth of the cell, was about $4.36 \times 10^8$, which is well into the turbulent regime. His key observations were:
\begin{enumerate}
\item The amplitude of temperature fluctuations was largest close to the base of the stable layer.
\item He released dye into the stable region, some of which was entrained into the convecting region to reveal the existence of elongated plume structures that extended from the base of the lower layer to the base of the stable layer.
\end{enumerate}
Townsend \cite{townsend1964} attributed the large amplitude of the temperature fluctuations to the generation of internal gravity waves in the stable layer. These waves were generated at the interface between stable and unstable regions by the random impingement of plumes originating at the bottom surface. A systematic measurement of the heat flux could not be made due to heat loss from the sidewalls.

Deardorff \emph{et al.} \cite{deardorff1969} took a different approach to study the dynamics of penetrative convection. Using water as the working fluid, and a temperature range far from the temperature of maximum density, their initial condition was one of stable stratification. Convection ensued once the temperature of the bottom plate was increased. The motivation of this configuration was to understand the lifting of the inversion layer due to heating of the ground, and thus the central focus was to understand the evolution of the convecting layer. Their theoretical model predicted that the thickness of the convecting layer grows diffusively ($\propto \sqrt{t}$, where $t$ is time) when the heat flux from the bottom plate was assumed to be constant. However, when a constant temperature was imposed at the bottom plate, they derived a modified evolution equation whose results were in agreement with measurements. The best fit to their theoretical solution gave the growth of the layer as $\sim t^{0.41}$ (Figure 11 of \cite{deardorff1969}). This indicates that the results for constant temperature and constant flux conditions are not substantially dissimilar, at least for the growth of the layer in this configuration. This is also supported by the fact that the heat transport in Rayleigh-B\'enard convection is the same for constant temperature and constant flux conditions \citep{doering2009}. Similar theoretical models have been constructed by Tennekes \cite{tennekes1973} and Mahrt and Lenschow \cite{mahrt1976} to study the evolution of the convective layer. The model of Mahrt and Lenschow \cite{mahrt1976} is obtained by integrating the equations of motion in the convecting layer, and it reduces to that of Tennekes \cite{tennekes1973} when shear generation by turbulence is neglected.

Penetrative convection is also important in the study of stars. A typical star is comprised of three regions: an inner radiative zone, an outer convective zone, and the tachocline, which is a transition layer between the radiative and convective zones, and is stably stratified \cite{spiegel1992}. Cold plumes from the outer convective zone penetrate the upper layers of the tachocline generating internal gravity waves, which are thought to play an important role in the turbulent transport of momentum in the tachocline \cite{brummell2002, dintrans2005, Lecoanet:2015}. Hence, a detailed study of penetrative convection is necessary for the understanding of the coupling between these different zones and the effects of that coupling on the magnetic field of the star.

The situation studied here bears resemblance to penetrative convection in an internally heated fluid, where the fluid, which is bound by horizontal surfaces maintained at equal temperatures, is non-uniformly \cite{whitehead1970} or uniformly \cite{goluskin2012, goluskin2016} heated. The presence of the heat source leads to the generation of an unstable upper layer and a stable bottom layer. The relevant questions for this setting are \cite{goluskin2016}: (1) how does the heat flux vary with the strength of the heat source? and (2) how does the mean temperature of the fluid vary with the strength of the heat source?  An important distinction from our work is that, due to the asymmetry introduced by the heat source, the heat flux at the top and bottom surfaces are not equal in the stationary state. Also, the dependence of the heat flux on the heat source differs in two and three dimensions \cite{goluskin2016}. Additionally, we note here that Chen and Whitehead \cite{chen1968} had previously used the idea of non-uniform heating of the fluid layer to study finite-amplitude motions in the classical Rayleigh-B\'enard convection.

By modeling the fluid density with a piecewise linear phenomenological equation of state, with the same fixed linear increase in the unstable region and an arbitrary variable linear decrease (characterized by a parameter $\mathcal{S} = \left[2^{-8}, 2^{8}\right]$) in the stable region, \citet{couston2017} numerically studied the flow in a similar geometry.  Because of the quantitative difference between our equation of state and their parameter $\mathcal{S}$, we are not in a position to make a quantitative comparison to our work.  However, we note that in this and a related study \cite{couston2017, Lecoanet:2015} they found (a) the convective region to be similar to that of the classical Rayleigh-B\'enard convection for $\mathcal{S} \ge 100$ and (b) gravity waves at the density interface, heuristically as do we under certain conditions.

Both the astrophysical and geophysical settings in which penetrative convection is important offer a wide range of complications, such as rotation, not part of our study.  However, in the spirit of the original paper of Veronis \cite{veronis1963}, we have found further basic fluid mechanical processes free from the ravages of such complications and these are of interest to study in their own right.


In this paper we consider penetrative convection in a fluid that has a density maximum at a temperature between two horizontal plates.
We derive an evolution equation for the thickness of the convecting layer by integrating the heat equation in the unstable layer and by using a form for the horizontally averaged temperature field based on our previous studies of turbulent Rayleigh-B\'enard convection \citep{TSW2015_EPL}. We then compare the theory with the results from high-resolution numerical simulations for large Rayleigh numbers. Finally, we discuss the effects of boundary conditions on the flow field and on the heat transport.  

\section{Equations of Motion}
\begin{figure}
\begin{centering}
\includegraphics[trim = 50 50 0 0, clip, width = \linewidth]{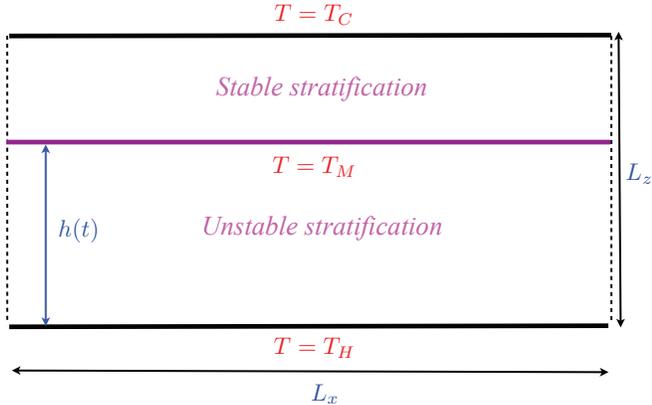} 
\caption{Schematic of the domain for penetrative convection. The purple line indicates the horizontal layer at which the density of the fluid is a maximum.}
\label{fig:PC}
\end{centering}
\end{figure}

Figure \ref{fig:PC} is a schematic of the system studied here. The width and depth of the domain are $L_x$ and $L_z$, respectively, the depth of the convecting layer is $h$, the bottom (top) plate is maintained at a temperature $T_H$  ($T_C$), and the fluid has a density maximum at a temperature $T_M$. The temperatures are such that $T_C < T_M < T_H$.

The fluid considered here is water, described well with the following equation of state \cite{veronis1963}:
\be
\rho = \rho_0 \, \left[1 - \alpha \, \left(T-T_M\right)^2 \right], 
\label{eqn:state}
\ee
showing that the fluid has a maximum density $\rho_0$ when $T=T_M$. Making the Boussinesq approximation, the equations of motion are
\be
\nabla \cdot \boldsymbol{u} = 0, 
\label{eqn:mass}
\ee
\be
\frac{\partial \boldsymbol{u}}{\partial t} +  \boldsymbol{u} \cdot \nabla  \boldsymbol{u} = -\frac{1}{\rho_0} \, \nabla p + g \, \alpha \, \left(T-T_M\right)^2 \, \boldsymbol{k} + \nu \, \nabla^2  \boldsymbol{u}, 
\label{eqn:NS}
\ee
and
\be
\frac{\partial T}{\partial t} +  \boldsymbol{u} \cdot \nabla  T = \kappa \, \nabla^2  T.
\label{eqn:heat}
\ee
Here, $\boldsymbol{u}(\boldsymbol{x},t)$ is the velocity field, 
$p(\boldsymbol{x},t)$ is the pressure field, $g$ is acceleration due to gravity, $\alpha$ is the coefficient of thermal expansion, $\boldsymbol{k}$ is the unit vector along the vertical, $\nu$ is kinematic viscosity, $T(\boldsymbol{x},t)$ is the temperature field, and $\kappa$ is the thermal diffusivity of the fluid. 

To non-dimensionalize Eqs. (\ref{eqn:mass}) -- (\ref{eqn:heat}), we choose $L_z$ as the length scale, $\Delta T = T_H-T_C$ as the temperature scale, $U_0 = \kappa/L_z$ as the velocity scale, $\rho_0 \, \nu \, \kappa/L_z$ as the pressure scale, and $t_0 = L_z^2/\kappa$ as the time scale. We also introduce the non-dimensional temperature $\theta$ as
\be
\theta = \frac{T-T_M}{\Delta T}.
\ee
Using these scales, but retaining the pre-scaled notation, save for the temperature field, we obtain
\be
\nabla \cdot \boldsymbol{u} = 0,
\label{eqn:mass1}
\ee
\be
\frac{\partial \boldsymbol{u}}{\partial t} +  \boldsymbol{u} \cdot \nabla  \boldsymbol{u} = Pr \left(-\nabla p + Ra \, \theta^2 \, \boldsymbol{k} +  \nabla^2  \boldsymbol{u} \right),
\label{eqn:NS1}
\ee
and
\be
\frac{\partial \theta}{\partial t} +  \boldsymbol{u} \cdot \nabla  \theta = \nabla^2  \theta, 
\label{eqn:heat1}
\ee
where
\be
Ra = \frac{g \, \alpha \, \left(\Delta T \right)^2 \, L_z^3}{\nu \, \kappa} \hspace{0.2cm} \text{and} \hspace{0.2cm} Pr = \frac{\nu}{\kappa}
\ee
are the Rayleigh and Prandtl numbers, respectively. Hence, in non-dimensional units we have $\theta (z = 0) = \theta_H$, $\theta_M = 0$, and $\theta(z=1) = \theta_C = - \theta_0$, where $\theta_0 >0$.

For velocity, the boundary conditions at the top and bottom surfaces are no-slip and no-penetration; and we assume periodicity in the horizontal direction.

\section{Numerical Scheme and Validation}

We use the Lattice Boltzmann Method \cite{benzi1992, chen1998, Succi2001, shan1997} to study penetrative convection for large Rayleigh number. The code developed has been extensively tested against results from spectral methods for shear and buoyancy driven flows \cite{TSW2015_iutam, TSW2015_EPL, TSW_PRL2017}. The buoyancy force is introduced into the lattice Boltzmann equation using the scheme of Guo \emph{et al.} \cite{guo2002}.

\begin{figure}
\begin{centering}
\includegraphics[trim = 0 0 0 0, clip, width = \linewidth]{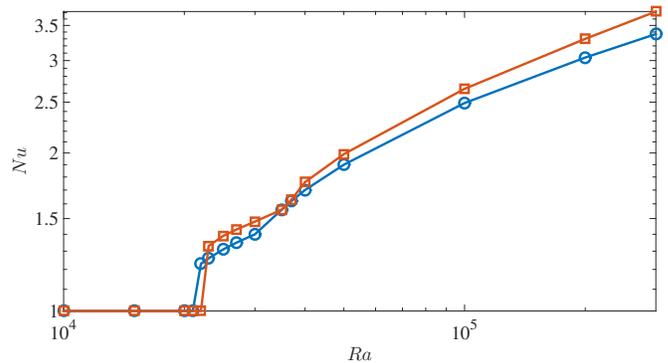} 
\caption{Comparison of $Nu(Ra)$ for $T_H = 8$ $^{\circ}$C, $T_M = 3.98$ $^{\circ}$C, $T_C = 0$ $^{\circ}$C, and $Pr = 11.6$ against the results of \cite{blake1984}. The squares denote values of \cite{blake1984} and the circles denote values from our simulations.}
\label{fig:blake}
\end{centering}
\end{figure}

The code has also been validated against the results of \citet{blake1984} for $\Gamma = L_x/L_z = 2$, $T_H = 8$ $^{\circ}$C, $T_M = 3.98$ $^{\circ}$C, $T_C = 0$ $^{\circ}$C, and $Pr = 11.6$. Figure \ref{fig:blake} shows the comparison of $Nu(Ra)$ with their simulations. Our values of $Nu$ are consistently lower than theirs, which we attribute  to the low resolution of $22 \times 42$ grid points used in their study; our resolution is an order of magnitude higher along both the horizontal and vertical directions.

We should note here that due to the presence of the stable layer, the time taken to reach a stationary state is much longer than in the classical Rayleigh-B\'enard setting. The steady state thickness of the convecting layer is reached when the conductive heat flux in the stable layer is equal to the heat flux from the unstable layer \cite{townsend1964, adrian1975}.

The results from numerical simulations presented in the following sections were obtained using $\Gamma \equiv L_x/L_z = 2$ and $Pr = 1$.

\section{Results}
\subsection{Analytical Results}
\subsubsection{Evolution of the Convecting Layer}
\begin{figure}
\begin{centering}
\includegraphics[trim = 0 50 0 0, clip, width = \linewidth]{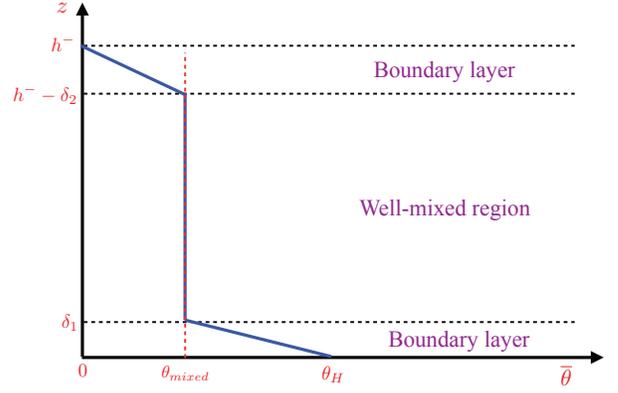} 
\caption{Figure shows the assumed profile for $\overline{\theta}(z,t)$.}
\label{fig:PC_theta}
\end{centering}
\end{figure}

Here, using Eq. (\ref{eqn:heat1}), we derive an evolution equation for the depth of the convecting layer, $h(t)$.  The flow is assumed incompressible, and thus Eq. (\ref{eqn:heat1}) can be written as
\be
\frac{\partial \theta}{\partial t} + \frac{\partial}{\partial x} \left(u \, \theta\right) + \frac{\partial}{\partial z} \left(w \, \theta\right) = \frac{\partial^2 \theta}{\partial x^2} + \frac{\partial^2 \theta}{\partial z^2}.
\ee
Integrating along $x$ and assuming periodicity, we find
\be
\frac{\partial \overline{\theta}}{\partial t} + \frac{\partial}{\partial z} \left(\overline{w' \, \theta'}\right) = \frac{\partial^2  \overline{\theta}}{\partial z^2},
\label{eqn:evolve1}
\ee
where
\be
\overline{\Psi} = \frac{1}{L_x} \, \int_0^{L_x} \! \Psi \, \mathrm{d}x
\ee
denotes the horizontal mean, and primes denote deviation from the horizontal means. Now, we integrate Eq. (\ref{eqn:evolve1}) along the vertical in the convecting region to find
\be
\int_0^{h^-} \! \frac{\partial \overline{\theta}}{\partial t} \, \mathrm{d}z = -\left.\left(\overline{w' \, \theta'} - \frac{\partial \overline{\theta}}{\partial z}\right)\right\vert_{z = h^-} + \left.\left(\overline{w' \, \theta'} - \frac{\partial \overline{\theta}}{\partial z}\right)\right\vert_{z = 0}.
\label{eqn:integ1}
\ee
Owing to the no-penetration condition at $z=0$, Eq. (\ref{eqn:integ1}) reduces to
\be
\int_0^{h^-} \! \frac{\partial \overline{\theta}}{\partial t} \, \mathrm{d}z = -\left.\left(\overline{w' \, \theta'} - \frac{\partial \overline{\theta}}{\partial z}\right)\right\vert_{z = h^-} - \left. \frac{\partial \overline{\theta}}{\partial z}\right\vert_{z = 0}.
\label{eqn:evolve2}
\ee
We assume that the dominant mode of heat transport in the stable layer is conduction, and by demanding the continuity of heat flux at the interface between the stable and unstable layers \cite[e.g.,][]{worster2004}, we have
\be
\left.\left(\overline{w' \, \theta'} - \frac{\partial \overline{\theta}}{\partial z}\right)\right\vert_{z = h^-} = -\left. \frac{\partial \overline{\theta}}{\partial z}\right\vert_{z = h^+}.
\label{eqn:evolve3}
\ee
Using condition (\ref{eqn:evolve3}) in Eq. (\ref{eqn:evolve2}), we find that
\be
\int_0^{h^-} \! \frac{\partial \overline{\theta}}{\partial t} \, \mathrm{d}z = \left. \frac{\partial \overline{\theta}}{\partial z}\right\vert_{z = h^+} - \left. \frac{\partial \overline{\theta}}{\partial z}\right\vert_{z = 0}.
\label{eqn:evolve5}
\ee
To evaluate the integral on the left hand side of Eq. (\ref{eqn:evolve5}), we make the following assumptions about $\overline{\theta}(z,t)$:
\begin{enumerate}
\item The convecting layer consists of a well-mixed region that is bounded by boundary layers on its top and bottom surfaces.
\item The small boundary-layer thicknesses ($\delta_1$ and $\delta_2$) are assumed to be constants. The argument being that the boundary layers reach a stationary state much more rapidly than the well-mixed region. 
\end{enumerate}
Figure \ref{fig:PC_theta} shows the assumed profile for $\overline{\theta}(z,t)$.
Based on this, we write
 \[ \overline{\theta}(z,t) =
  \begin{cases}
    \overline{\theta}_1 = \left(\theta_{\text{mixed}}-\theta_H\right) \, \frac{z}{\delta_1} + \theta_H;        & \quad \text{if } 0 \le z \le \delta_1,\\
    \overline{\theta}_2 = \theta_{\text{mixed}};  & \quad \text{if } \delta_1 \le z \le h^--\delta_2,\\
    \overline{\theta}_3 = \left(\frac{h^--z}{\delta_2}\right) \, \theta_{\text{mixed}};  & \quad \text{if } h^--\delta_2 \le z \le h^-.
  \end{cases}
\]
The integral in Eq. (\ref{eqn:evolve5}) can now be written as
\be
\int_0^{h^-} \! \frac{\partial \overline{\theta}}{\partial t} \, \mathrm{d}z = \int_0^{\delta_1}  \frac{\partial \overline{\theta}_1}{\partial t} \, \mathrm{d}z + \int_{\delta_1}^{h^--\delta_2} \frac{\partial \overline{\theta}_2}{\partial t} \, \mathrm{d}z + \int_{h^--\delta_2}^{h^-} \frac{\partial \overline{\theta}_3}{\partial t} \, \mathrm{d}z.
\ee
Assuming $\theta_{\text{mixed}}$ to be a constant, the integrals are easily evaluated to yield
\be
\int_0^{h^-} \! \frac{\partial \overline{\theta}}{\partial t} \, \mathrm{d}z = \theta_{\text{mixed}} \, \frac{\mathrm{d}h}{\mathrm{d}t},
\ee
and hence, Eq. (\ref{eqn:evolve5}) becomes
\be
\theta_{\text{mixed}} \, \frac{\mathrm{d}h}{\mathrm{d}t} = \left. \frac{\partial \overline{\theta}}{\partial z}\right\vert_{z = h^+} - \left. \frac{\partial \overline{\theta}}{\partial z}\right\vert_{z = 0}.
\ee
Moreover, we have
\be
\left. \frac{\partial \overline{\theta}}{\partial z}\right\vert_{z = h^+} = \frac{\theta_C - \theta_M}{1-h} = -\frac{\theta_0}{1-h},
\ee
and
\begin{widetext}
\be
- \left. \frac{\partial \overline{\theta}}{\partial z}\right\vert_{z = 0} = -\left.\frac{\partial \overline{T}}{\partial z}\right\vert_{z = 0} \, \left(\frac{\Delta T}{L_z}\right)^{-1} = -\left.\frac{\partial \overline{T}}{\partial z}\right\vert_{z = 0}  \, \left(\frac{\Delta T_1}{h} \frac{h}{L_z} \frac{\Delta T}{\Delta T_1}\right)^{-1} = \frac{Q}{h} \, \theta_H, 
\ee
\end{widetext}
where $\Delta T_1 = T_H-T_M$ and $Q$ ($>0$) is the non-dimensional heat flux delivered to the convecting region. Hence, we have the following evolution equation for the thickness of the convecting region,
\be
\frac{\mathrm{d}h}{\mathrm{d}t} = -\frac{\Lambda}{\gamma \, \left(1-h\right)} + \frac{1}{\gamma} \, \frac{Q}{h},
\label{eqn:thickness_evolve}
\ee
where $\theta_{\text{mixed}} = \gamma \, \theta_H$, with $0 < \gamma <1$ a constant, and $\Lambda = \theta_0/\theta_{H}$.  We note that in our approach the evolution equation has been obtained by assuming a profile for the mean temperature based on our quantitative understanding of the flow structure in classical turbulent Rayleigh-B\'enard convection. Any need to parametrize the turbulent heat flux is circumvented by the requirement that the heat flux be continuous at the interface between the stable and unstable layers. 

Our analysis also reveals that, in addition to $Ra$ and $Pr$, there is another governing parameter in the system, which is given by 
\be
\Lambda = \frac{\theta_M - \theta_C}{\theta_H - \theta_M} = \frac{\theta_0}{\theta_H}.
\ee
In general, the range of values $\Lambda$ can take is $[0, \infty)$. A large (small) value of $\Lambda$ indicates that the stable layer is strongly (weakly) stratified. The characteristics of the flow depend sensitively on the value of $\Lambda$, and hence this is a very important parameter in the description of penetrative convection.

In Eq. (\ref{eqn:thickness_evolve}), there are different balances between the terms for different times, which is heuristically like the balances found in double-diffusive \cite{worster2004} and solidification problems \cite{turner1986, huppert1992}. Let $\mathcal{T}_t$ be the time at which the initial transients decay and $\mathcal{T}_g$ be the time beyond which the flow reaches a stationary state. The convective layer evolves in the following three stages:
\begin{enumerate}

\item \underline{Transient state: $0 \le t \le \mathcal{T}_i$} \\
The dominant balance during this time period is:
\be
\frac{\mathrm{d}h}{\mathrm{d}t} = -\frac{\Lambda}{\gamma \, \left(1-h\right)},
\ee
which implies that the convective layer shrinks. This is expected on the grounds that the flow responds to the bottom heat flux on a time scale of $\mathcal{O}(\mathcal{T}_i)$,  during which the second term of Eq. (\ref{eqn:thickness_evolve}) is smaller. The value of $\mathcal{T}_i$ would depend on $Ra$ and $\Lambda$, and in general can be expected to decrease with increasing $Ra$ and decreasing $\Lambda$.

\item \underline{Growth: $\mathcal{T}_i < t \le \mathcal{T}_g$} \\
During this stage we have
\be
\frac{\mathrm{d}h}{\mathrm{d}t} = \frac{1}{\gamma} \, \frac{Q}{h},
\label{eqn:growth}
\ee
which implies that the thickness of the layer increases with time. For $Q$ constant, the solution to Eq. (\ref{eqn:growth}) is
\be
h(t) = \sqrt{h_0^2 + \left(\frac{2 \, Q}{\gamma}\right) \, t},
\ee
where $h_0$ is the thickness at $t=0$. Thus, our analysis recovers the result discussed above that the convective layer grows diffusively for constant heat flux when $Ra \gg 1$  \cite{deardorff1969, tennekes1973, farmer1975}. 

\item \underline{Steady state: $t > \mathcal{T}_g$} \\
In the final stage, the flow reaches a steady state and Eq. (\ref{eqn:thickness_evolve}) becomes
\be
\frac{\Lambda}{\gamma \, \left(1-h_s\right)} = \frac{1}{\gamma} \, \frac{Q}{h_s},
\ee
where $h_s$ is final thickness of the layer, which is
\be
h_s = \frac{Q}{\Lambda + Q}.
\label{eqn:growth1}
\ee
For a fixed $Q$, when the upper layer is unstratified ($\Lambda = 0$), Eq. (\ref{eqn:growth1}) gives $h_s = 1$ and the convective layer occupies the whole domain. In the opposite limit of very strong stratification of the upper layer ($\Lambda \rightarrow \infty$), we have $h_s \rightarrow 0$. Both of these limits are found in our simulations.

\end{enumerate}

We note that the expression for effective $Nu$ ($Q$ in our notation) in the work of Moore \& Weiss (Eq. (15) of \cite{moore1973}) reduces to Eq. (\ref{eqn:growth1}) after some algebraic manipulation.

\subsection{Numerical Results}
\subsubsection{Thickness of the Convecting Layer}
\begin{figure}
\begin{centering}
\includegraphics[trim = 0 0 0 0, clip, width = \linewidth]{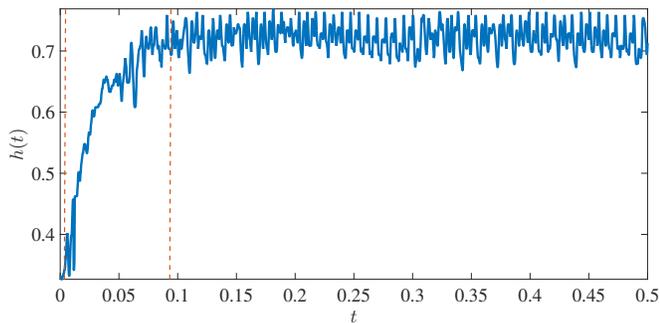} 
\caption{Evolution of the thickness of the convecting layer for $Ra = 10^7$ and $\Lambda = 2$. The dashed lines separate the three stages of evolution, as discussed in the main text.}
\label{fig:thick1}
\end{centering}
\end{figure}

\begin{figure}
\begin{centering}
\includegraphics[trim = 0 0 0 0, clip, width = \linewidth]{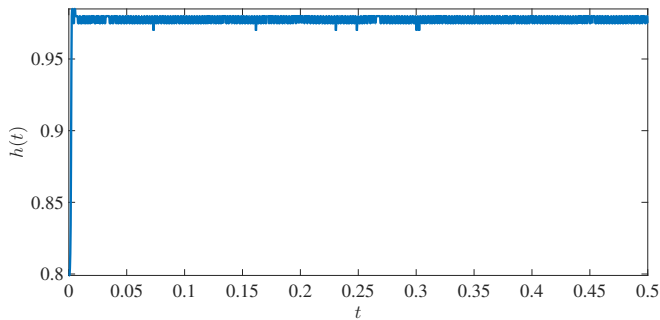} 
\caption{Evolution of the thickness of the convecting layer for $Ra = 10^7$ and $\Lambda = 0.25$.}
\label{fig:thick2}
\end{centering}
\end{figure}

We compute the thickness of the convecting layer, $h(t)$, which is defined as the height at which $\overline{\theta} = 0$. Figures \ref{fig:thick1} and \ref{fig:thick2} show the evolution of the convecting layer for $Ra = 10^7$ and $\Lambda = 2$ and $0.25$, respectively.
A fit to the region where $h(t)$ increases in time for $\Lambda = 2$ in Figure \ref{fig:thick1} gives $h(t) \propto t^{0.23}$; whereas, for $\Lambda = 0.25$ one obtains $h(t) \propto t$. This shows that the growth for the convecting layer is much faster when $\Lambda$ is small, which arises from two effects.  Firstly, the initial thickness of the convecting layer is larger for $\Lambda = 0.25$ than for $\Lambda = 2$ (see Figures \ref{fig:thick1} and \ref{fig:thick2} for thickness at $t=0$). The convective motions are more vigorous in the former case, leading to faster growth.  Secondly, for lower values of $\Lambda$, the developing convecting layer experiences little resistance in entraining fluid from the stable layer, which again leads to a faster growth.

\begin{figure}
\begin{centering}
\includegraphics[trim = 0 0 0 0, clip, width = \linewidth]{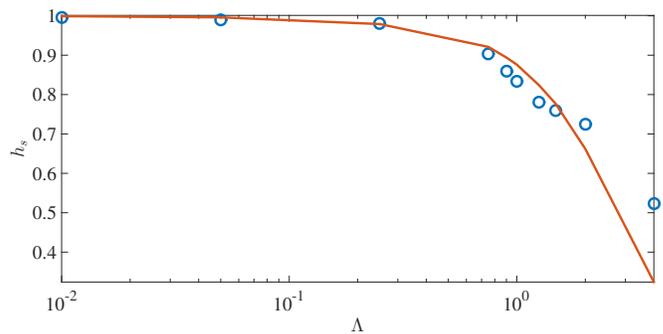} 
\caption{The averaged thickness, $h_s$, vs. $\Lambda$ for $Ra = 10^7$. Circles are the values obtained from simulations, and the solid line is from the theory. }
\label{fig:thickness_lambda4}
\end{centering}
\end{figure}

Once the flow has reached a stationary state, we compute the averaged thickness, $h_s$.  Figure \ref{fig:thickness_lambda4} shows $h_s$ as a function of $\Lambda$ for $Ra=10^7$. The agreement between the theory and simulations is very good for small $\Lambda$, but decreases for large $\Lambda$.  The large  $\Lambda$ behavior arises from the suppression of convective motions, and hence mixing, in the interior of the unstable region.  This leads to both conduction and convection becoming important throughout the unstable layer, and hence the temperature profile assumed in the theoretical analysis is no longer valid for these large $\Lambda$.

\subsubsection{Temperature Field}

\begin{figure*}
\begin{centering}
\includegraphics[trim = 95 0 0 0, clip, scale=0.375]{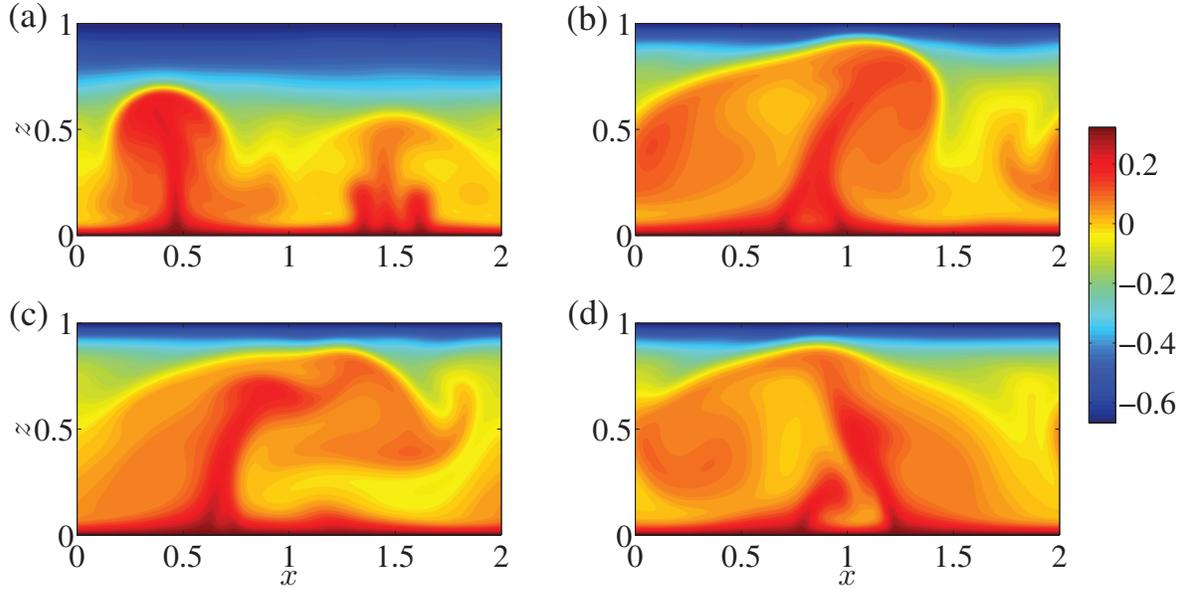} 
\caption{Temperature field for $Ra = 10^7$ and $\Lambda = 2$ at different times: (a) $t = 0.016$; (b) $t = 0.079$; (c) $t = 0.16$; and (d) $t = 0.32$. The structure of the flow here is in qualitative agreement with the experiments of Townsend \cite{townsend1964} with $\Lambda \approx 5$. See supplemental movie.}
\label{fig:temp_theta2}
\end{centering}
\end{figure*}

\begin{figure*}
\begin{centering}
\includegraphics[trim = 95 0 0 0, clip, scale=0.375]{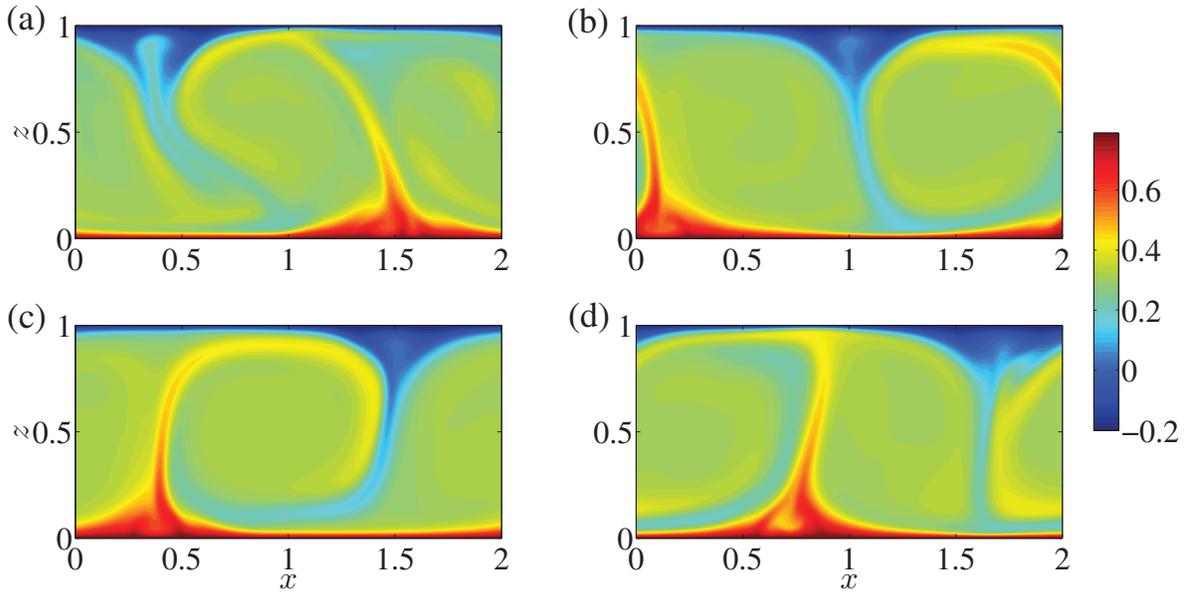} 
\caption{Temperature field for $Ra = 10^7$ and $\Lambda = 0.25$ at different times: (a) $t = 0.016$; (b) $t = 0.079$; (c) $t = 0.16$; and (d) $t = 0.32$. See supplemental movie.}
\label{fig:temp_theta025}
\end{centering}
\end{figure*}

Figures \ref{fig:temp_theta2} and \ref{fig:temp_theta025} show the time evolution of the temperature field for $Ra=10^7$ and $\Lambda = 2$ and $0.25$. These values of $\Lambda$ were chosen to clearly reveal the effects of the stable layer stratification on the flow characteristics. 
In Figure \ref{fig:temp_theta2}, the plumes that are generated from the hot bottom plate do not penetrate the stable layer because the strength of the stratification. The fluid from the stable layer is entrained slowly, and the flow takes a very long time to reach a stationary state. This is also clearly seen in Figure \ref{fig:thick1}, where the growth of the convecting layer is subdiffusive. Additionally, we observe internal gravity waves generated at the interface between the stable and unstable layers, as well as in the interior of the stable layer in Figures \ref{fig:temp_theta2}(b) -- \ref{fig:temp_theta2}(d). The structure of the flow here is in qualitative agreement with the experimental observations of Townsend \cite{townsend1964}, with $\Lambda \approx 5$.

In contrast to this, the plumes penetrate the stable layer when $\Lambda = 0.25$. In fact, the temperature fields closely resemble those in the Rayleigh-B\'enard problem, where the fluid has a linear equation of state.  Hence we intuitively expect that the Rayleigh-B\'enard problem is realized in the limit of $\Lambda \rightarrow 0$. Indeed, the rapid growth of the convecting layer, as seen in Figure \ref{fig:thick2}, is partly due to the weak stratification of the stable layer.

\begin{figure}
\begin{centering}
\includegraphics[trim = 0 0 0 0, clip, width = \linewidth]{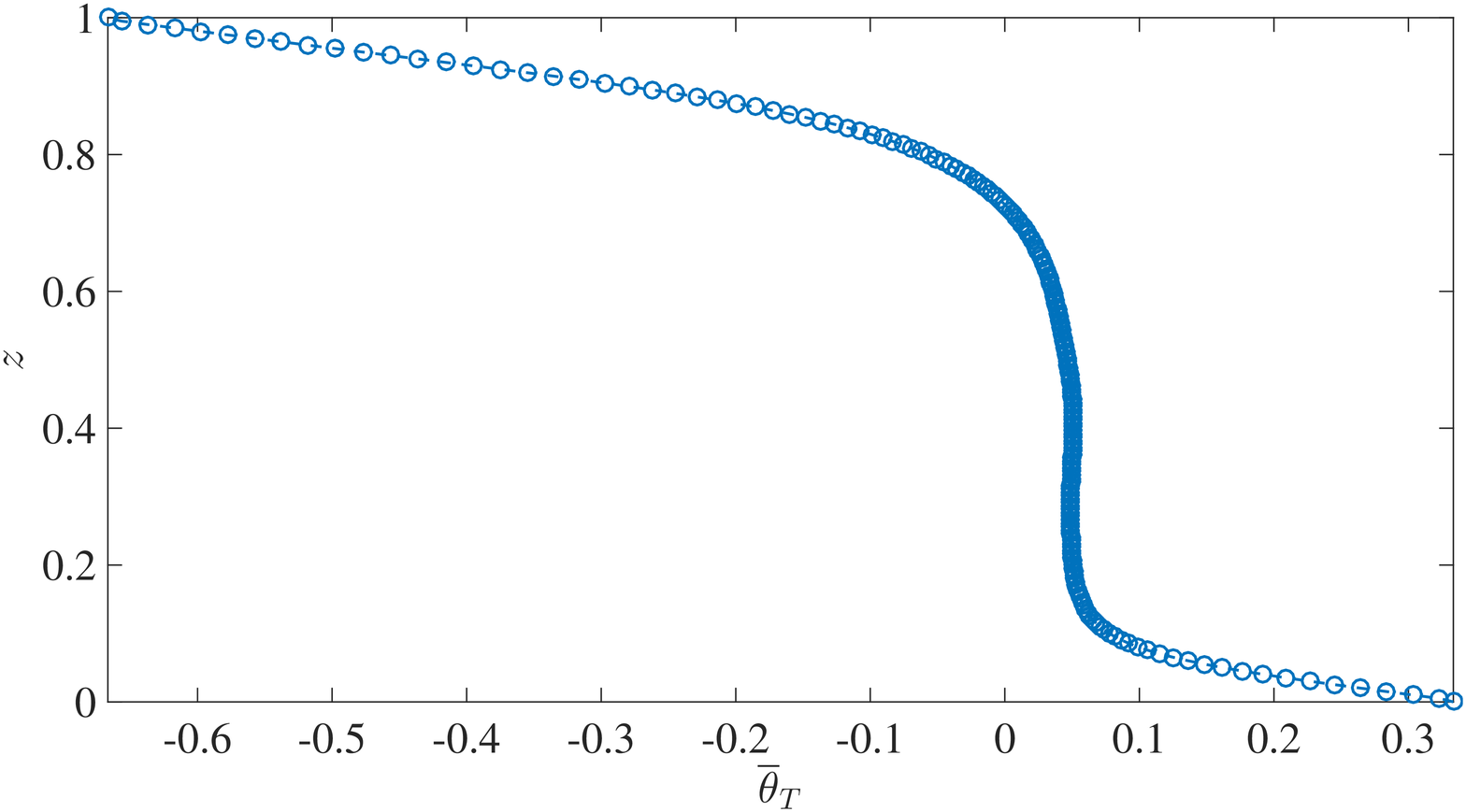} 
\caption{Mean temperature profile for $Ra=10^7$ and $\Lambda = 2$.}
\label{fig:profile1}
\end{centering}
\end{figure}

\begin{figure}
\begin{centering}
\includegraphics[trim = 0 0 0 0, clip, width = \linewidth]{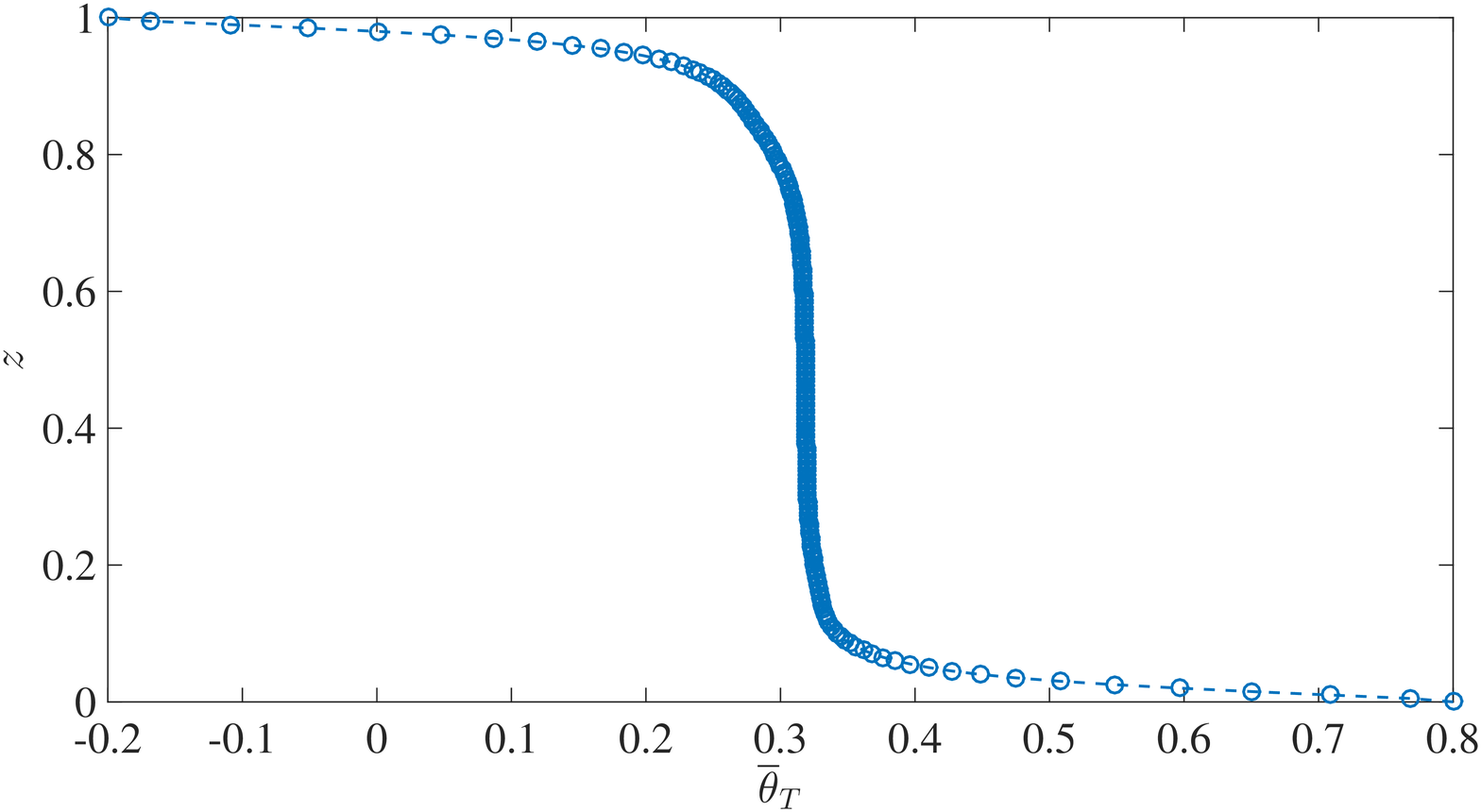} 
\caption{Mean temperature profile for $Ra=10^7$ and $\Lambda = 0.25$.}
\label{fig:profile2}
\end{centering}
\end{figure}

\begin{figure}
\begin{centering}
\includegraphics[trim = 0 0 0 0, clip, width = \linewidth]{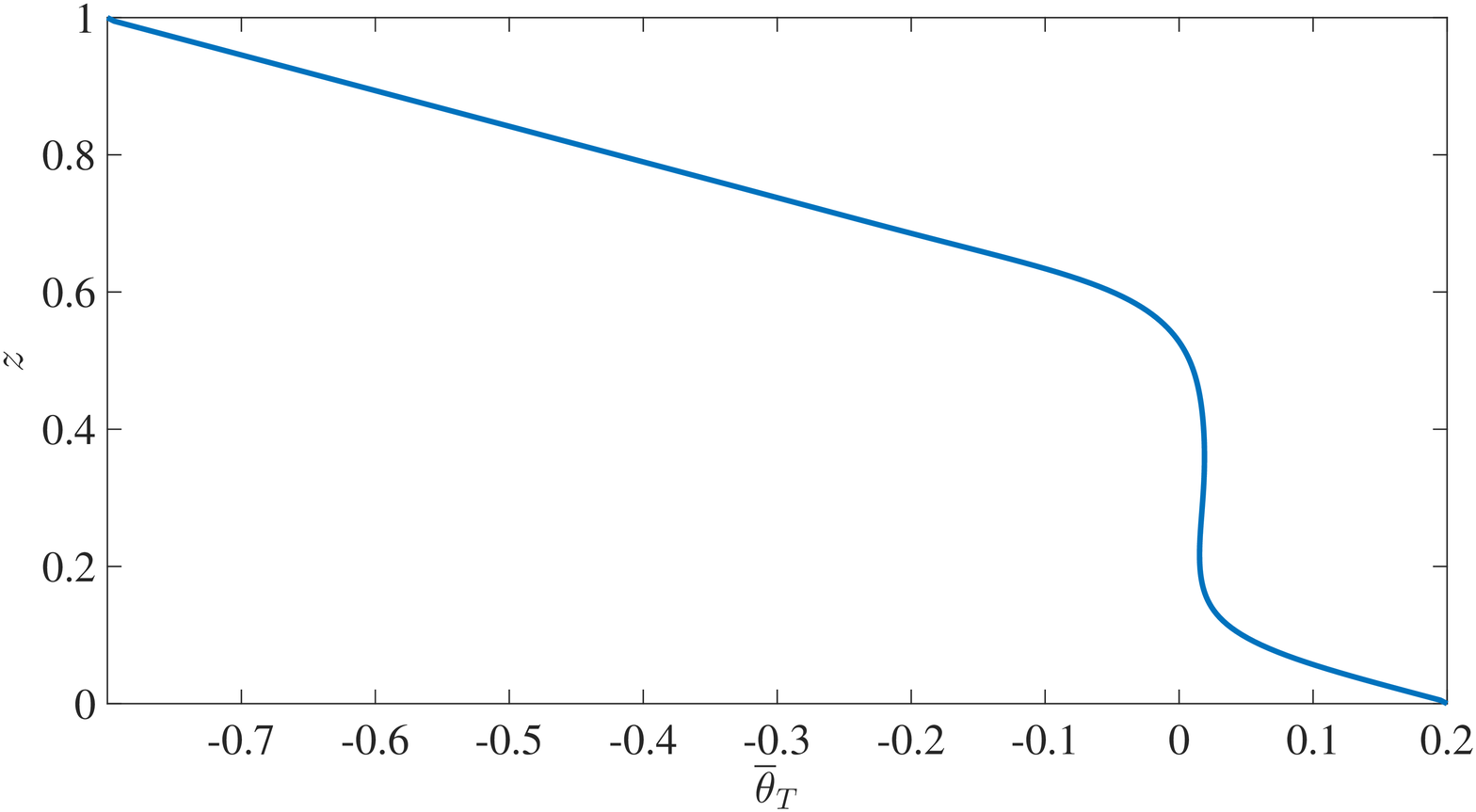} 
\caption{Mean temperature profile for $Ra=10^7$ and $\Lambda = 4$. This temperature profile is in qualitative agreement with the measurements of Adrian \cite{adrian1975}, who had $\Lambda \approx 5$ -- $6$.}
\label{fig:profile3}
\end{centering}
\end{figure}

The effects of $\Lambda$ can also be discerned by studying the temporally and horizontally averaged temperature profiles. To that end, figures \ref{fig:profile1} and \ref{fig:profile2} show $\overline{\theta}_T(z)$ for $Ra=10^7$ and $\Lambda = 2$ and $\Lambda = 0.25$, respectively. The temperature profile for $\Lambda = 2$ is more asymmetric than for $\Lambda = 0.25$. The stable layer is much thicker for $\Lambda = 2$, which is seen by the linear profile extending from $z=1$ to $z=0.8$. However, for $\Lambda = 0.25$ the top/bottom symmetry of the temperature profile closely resembles that from turbulent Rayleigh-B\'enard convection, consistent with the argument that the penetrative convective flow approaches that of the classical Rayleigh-B\'enard problem as $\Lambda \rightarrow 0$.
Additionally, Figure \ref{fig:profile3}  shows the averaged temperature profile for $Ra=10^7$ and $\Lambda = 4$, which is in qualitative agreement with the experiments of Adrian \cite{adrian1975} who had $\Lambda \approx 5$ -- $6$.

\subsubsection{Metastability of Plume Patterns}
Another interesting consequence of the presence of the stable layer is its effect on the dynamics of plume generation in the convective layer. In Rayleigh-B\'enard convection, the flow, for a given $Ra$ and $\Gamma$, settles into a stationary state with a fixed number of convection rolls that transport heat from the bottom wall to the top \cite{doering2009}. However, in penetrative convection, we find that for large $\Lambda$ and certain $Ra$, the flow structures enter a metastable state.

\begin{figure*}
\begin{centering}
\includegraphics[trim = 95 0 0 0, clip, scale=0.375]{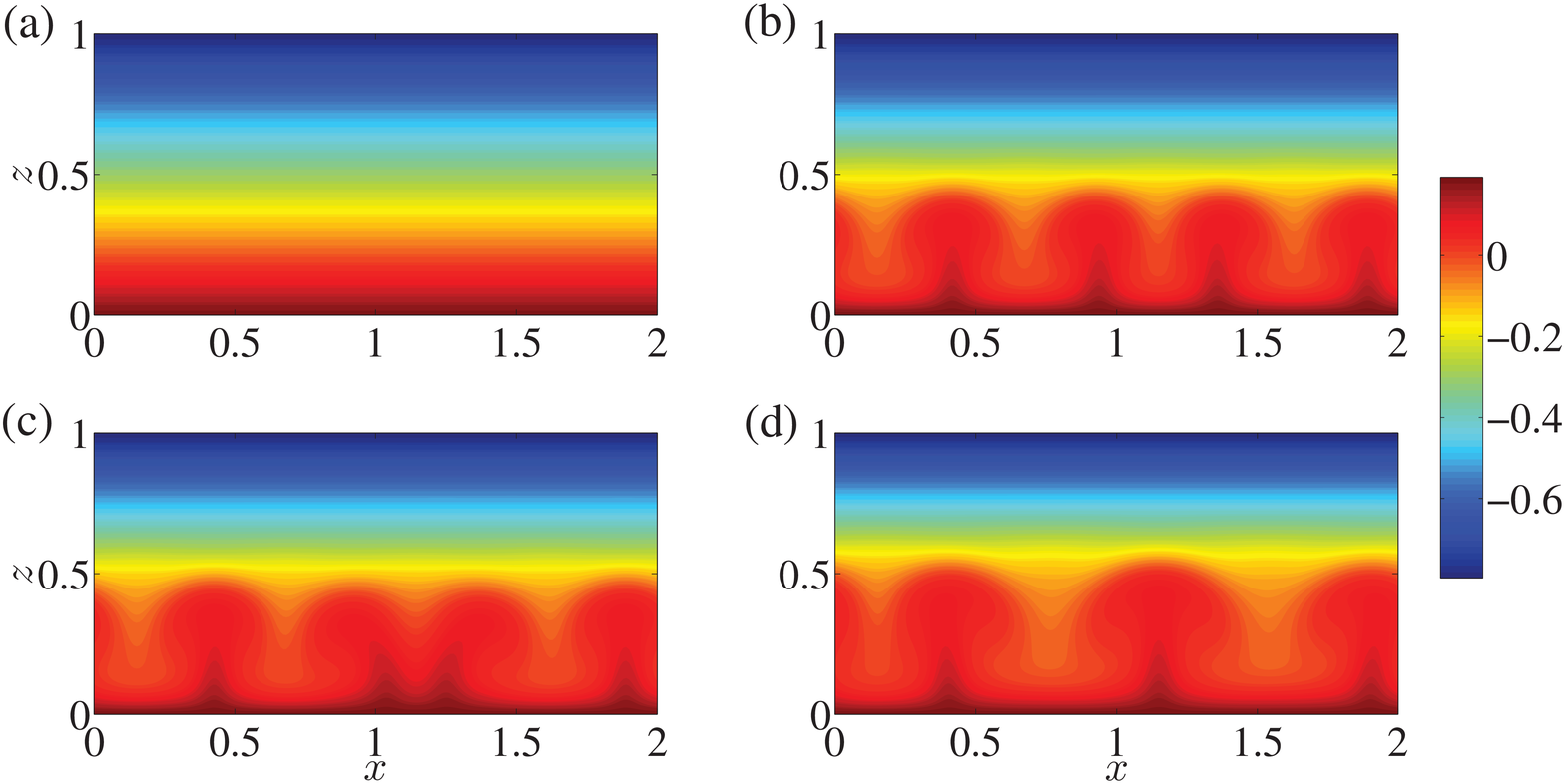} 
\caption{Evolution of temperature field for $Ra = 5 \times 10^6$ and $\Lambda = 4$ at times: (a) $t = 0$; (b) $t = 0.07$; (c) $t = 0.09$; and (d) $t = 0.12$. See supplemental movie.}
\label{fig:L4_1}
\end{centering}
\end{figure*}

\begin{figure*}
\begin{centering}
\includegraphics[trim = 95 0 0 0, clip, scale=0.375]{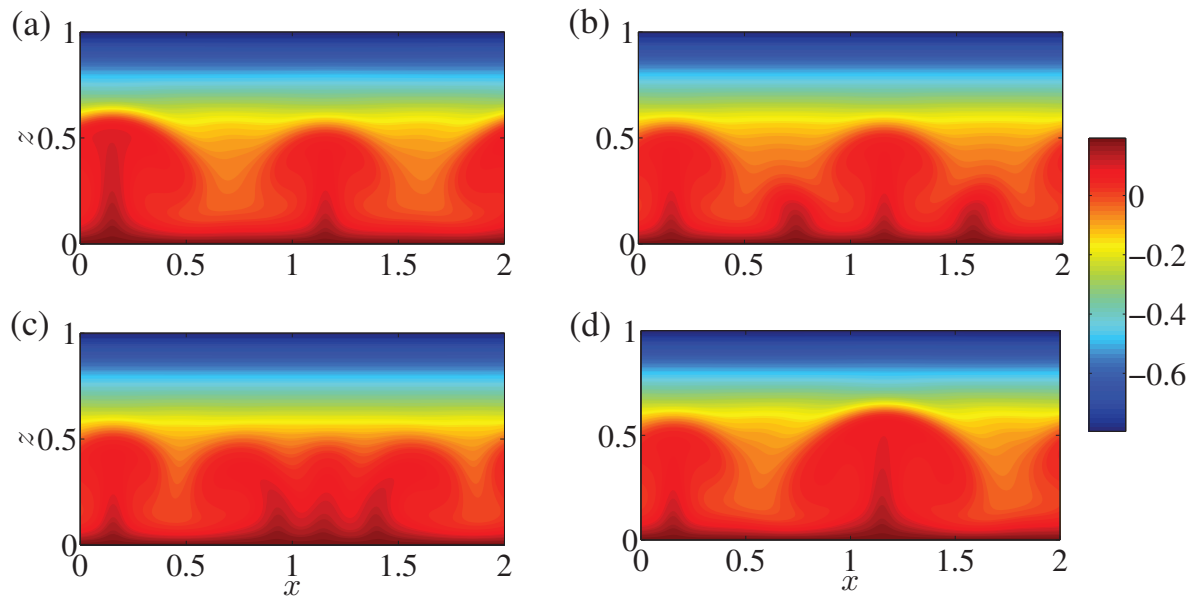} 
\caption{Evolution of temperature field for $Ra = 5 \times 10^6$ and $\Lambda = 4$ at times: (a) $t = 0.16$; (b) $t = 0.21$; (c) $t = 0.22$; and (d) $t = 0.24$. See supplemental movie.}
\label{fig:L4_2}
\end{centering}
\end{figure*}

Figures \ref{fig:L4_1} and \ref{fig:L4_2} show the evolution of temperature field for $Ra = 5 \times 10^6$ and $\Lambda = 4$. Focussing on the number of upwelling plumes, we see that there are four plumes in Figure \ref{fig:L4_1}(b) and as $h$ increases this configuration becomes unstable and two of the four plumes merge in Figure \ref{fig:L4_1}(c) forming now a total of three plumes as seen in Figure \ref{fig:L4_1}(d).

As $h$ increases further, the new configuration becomes unstable and two of the three plumes merge, giving rise to a total of two plumes [Figure \ref{fig:L4_2}(a)]. With increasing time, two smaller plumes are generated which then merge with one of the two larger plumes [Figures \ref{fig:L4_2}(a) - \ref{fig:L4_2}(d)]. This cycle of generation and merger of plumes continues and the flow does not settle into a stationary state with respect to flow structure.
It is interesting to note that qualitatively similar observations of plume merger and generation were made by \citet{whitehead1970} in their study of penetrative convection in internally heated fluid.

\begin{figure}
\begin{centering}
\includegraphics[trim = 0 0 0 0, clip, width = \linewidth]{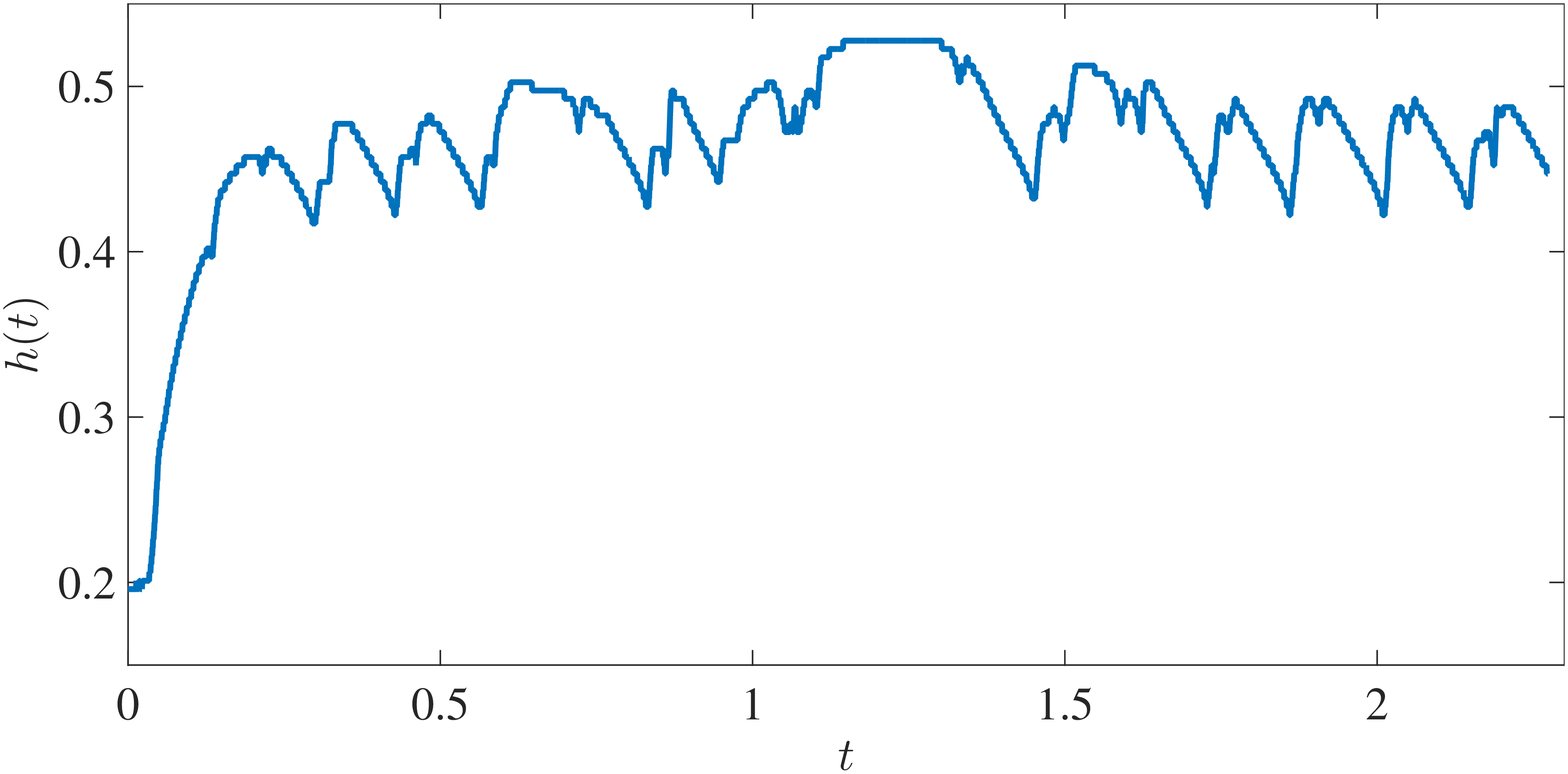} 
\caption{Evolution of the thickness of convective layer for $Ra = 5 \times 10^6$ and $\Lambda = 4$. The low-frequency oscillations are due to the impingement of plumes after merger events. See supplemental movie.}
\label{fig:L4_thickness}
\end{centering}
\end{figure}

These merger events give rise to more energetic plumes that then impinge upon the stable layer. This is seen in Figure \ref{fig:L4_2}(d). However, because of the stability of the upper layer, the plumes only generate low-frequency oscillations, as seen in Figure \ref{fig:L4_thickness}.

\subsubsection{Heat Transport}
The non-dimensional heat flux, $Nu$, from the lower surface to the upper surface can be obtained using
\be
Nu = - \left. \left(\frac{\partial \overline{\theta}_T}{\partial z}\right)_T \right|_{z=0}.
\label{eqn:Nu}
\ee
We note here that only the choice of $L_z$ as the characteristic length scale and $\Delta T = T_H - T_C$ as the characteristic temperature scale gives $Nu = 1$ when $Ra=0$. The simulations were run for sufficiently long times to obtain converged statistics to compute $Nu$. 

\begin{figure}
\begin{centering}
\includegraphics[trim = 50 70 20 125, clip, width = \linewidth]{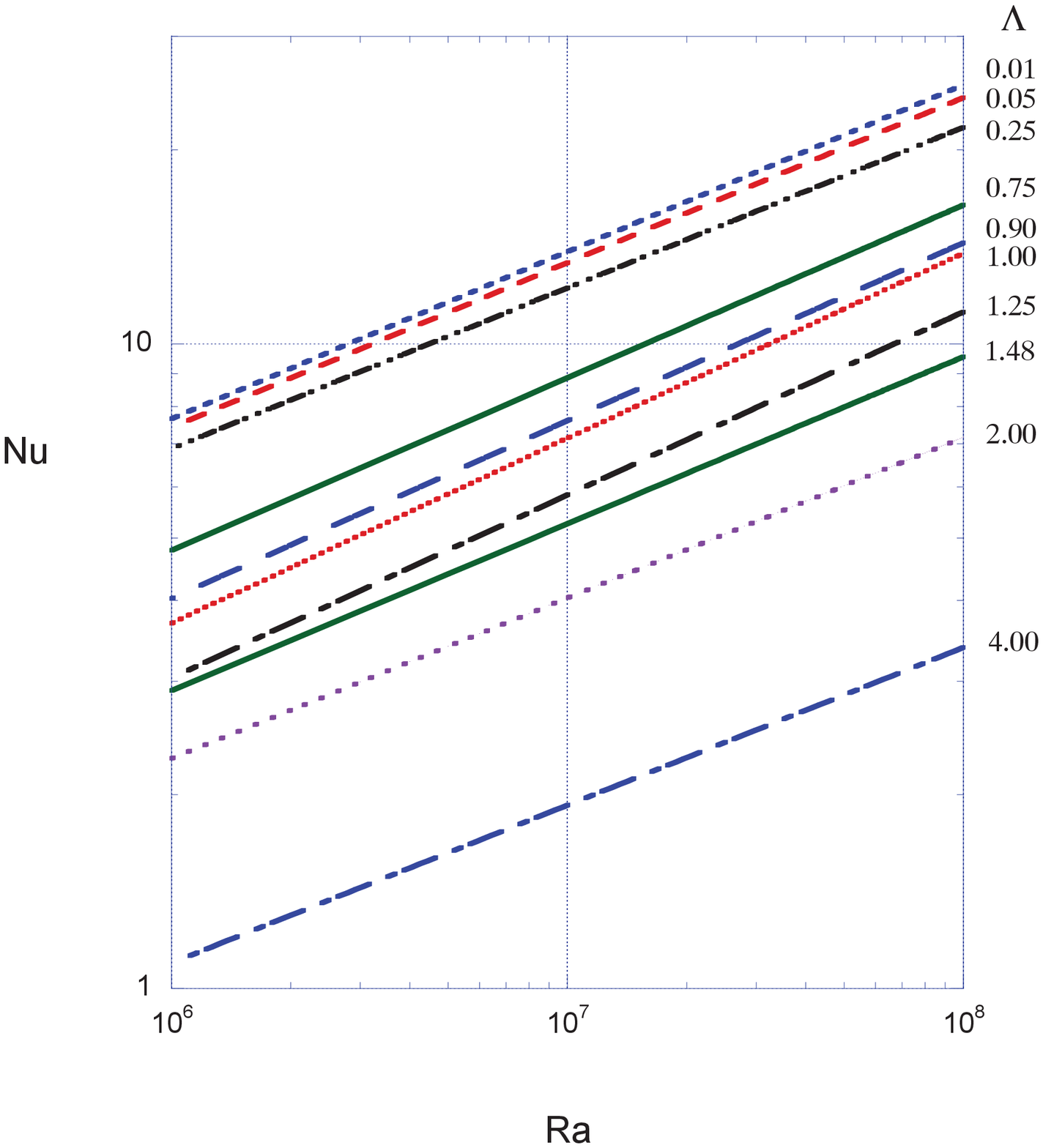} 
\caption{Power-law fits for $Nu$($Ra$, $\Lambda$) over $Ra = \left[10^6, 10^8\right]$ and $\Lambda = \left[0.01, 4\right]$. The individual power-laws are: 
(1) $\Lambda = 0.01$: $Nu = 0.214 \times Ra^{0.259}$; (2) $\Lambda = 0.05$: $Nu = 0.216 \times Ra^{0.256}$; (3) $\Lambda = 0.25$: $Nu = 0.221 \times Ra^{0.249}$; (4) $\Lambda = 0.75$: $Nu = 0.118 \times Ra^{0.268}$; (5) $\Lambda = 0.90$: $Nu = 0.089 \times Ra^{0.276}$; (6) $\Lambda = 1.00$: $Nu = 0.07 \times Ra^{0.287}$; (7) $\Lambda = 1.25$: $Nu = 0.06 \times Ra^{0.284}$; (8) $\Lambda = 1.48$: $Nu = 0.081 \times Ra^{0.259}$; (9) $\Lambda = 2.00$: $Nu = 0.073 \times Ra^{0.249}$; and (10) $\Lambda = 4.00$: $Nu = 0.037 \times Ra^{0.245}$.}
\label{fig:Nu1}
\end{centering}
\end{figure}

Figure \ref{fig:Nu1} shows the least-squares fits for $Nu$($Ra$, $\Lambda$) data for $Ra = \left[10^6, 10^8\right]$ and $\Lambda = \left[0.01, 4\right]$. For each $\Lambda$, the relation between $Nu$ and $Ra$ is sought in terms of a power law: $Nu = A \times Ra^{\beta}$.  Clearly, for a fixed value of $Ra$, $Nu$ increases with decreasing $\Lambda$. This is due to the fact the stability of the upper layer decreases as $\Lambda$ decreases, which in turn leads to more vigorous convective motions in the unstable layer and larger heat transport. For $\Lambda = 4$, there is no appreciable convective motion even when $Ra = 10^6$ and the heat transport in the entire domain is dominated by conduction.

\begin{figure}
\begin{centering}
\includegraphics[trim = 0 0 0 0, clip, width = \linewidth]{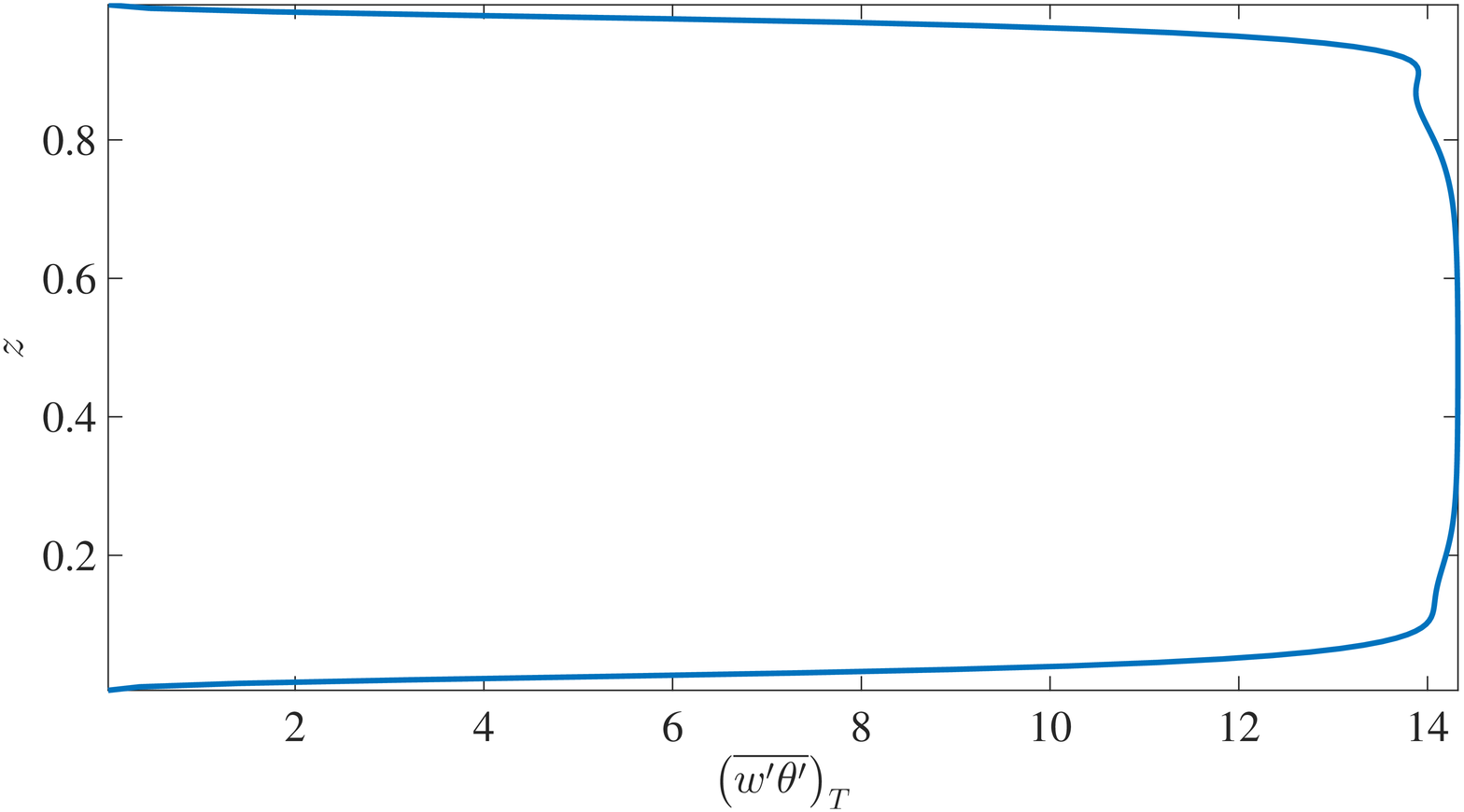} 
\caption{Variation of the convective heat flux, $Q_c$, with height for $Ra = 10^7$ and $\Lambda = 0.01$.}
\label{fig:HF001}
\end{centering}
\end{figure}

\begin{figure}
\begin{centering}
\includegraphics[trim = 0 0 0 0, clip, width = \linewidth]{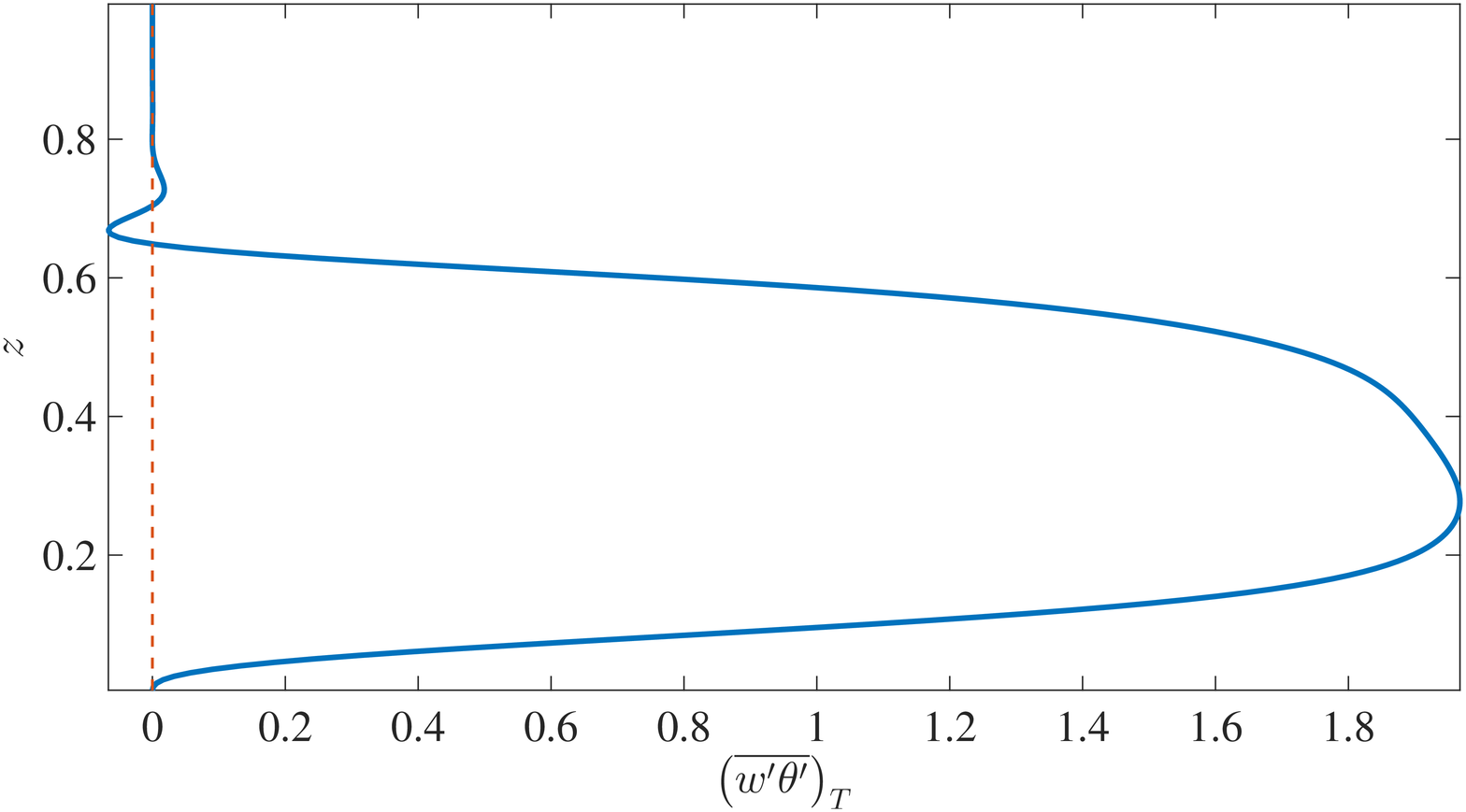} 
\caption{Variation of the convective heat flux, $Q_c$, with height for $Ra = 10^7$ and $\Lambda = 4$. The dashed vertical line is included to discern the change in sign in $Q_c$.}
\label{fig:HF4}
\end{centering}
\end{figure}

\begin{figure}
\begin{centering}
\includegraphics[trim = 0 0 0 0, clip, width = \linewidth]{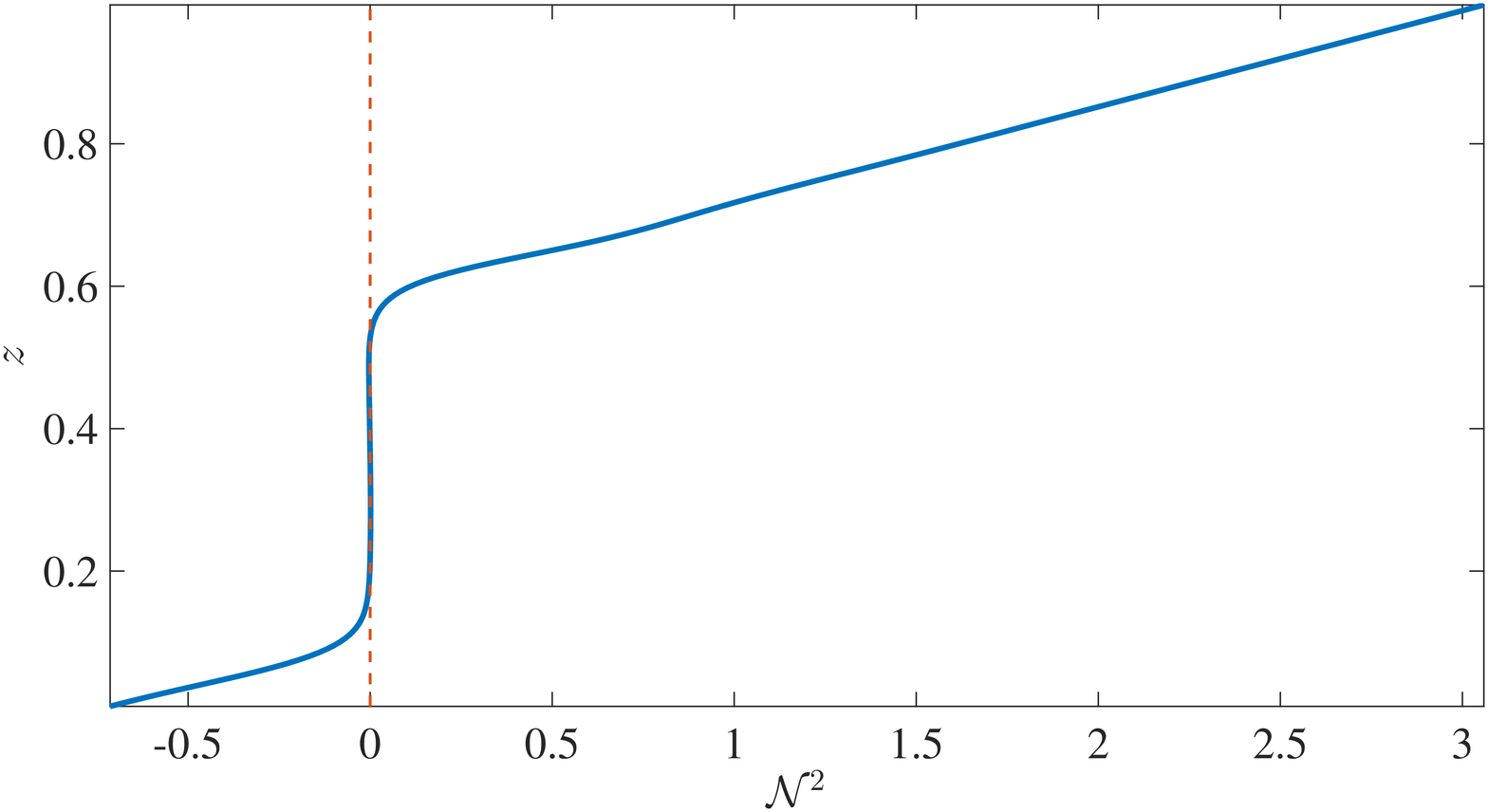} 
\caption{Variation of the Brunt-V\"{a}is\"{a}l\"{a} frequency, $\mathcal{N}^2$, with height for $Ra = 10^7$ and $\Lambda = 4$. The dashed vertical line is included to discern the change in sign in $\mathcal{N}^2$.}
\label{fig:BV_Lambda4}
\end{centering}
\end{figure}

Another quantity that is of interest is the convective heat flux, $Q_c = \left(\overline{w' \, \theta'}\right)_T$, and its variation with height. Deardorff \emph{et al.} \cite{deardorff1969} found that $Q_c$ remains positive in the convective region, but becomes negative near the interface due to entrainment of the fluid from the stable layer. Similar observations have also been made by Adrian \cite{adrian1975}. For $Ra = 10^7$, Figures \ref{fig:HF001} and \ref{fig:HF4} show how $Q_c$ changes as $\Lambda$ changes from 4 to 0.01, respectively.
It is clear that, for $\Lambda = 0.01$, except for the boundary layers, $Q_c$ is constant in the unstable region. Hence, in this case, convective motions transport nearly all the heat. On the other hand, when $\Lambda = 4$ convection is not the dominant mode of transport, even in the unstable region. This is reflected by the fact that $Q_c$ attains a maximum value, equal to $Nu$ for this case, in only a small region of the flow. Moreover, $Q_c$ changes sign again in the stable layer, which is due to the combined effects of entrainment of the fluid from the stable layer and the excitation of internal gravity waves. This is quantified by studying the height dependence of the Brunt-V\"{a}is\"{a}l\"{a} frequency \cite{turner1979}, which is defined in dimensional units as
\begin{equation}
\mathcal{N}^2 = - \, \frac{g}{\rho_0} \, \frac{\partial \rho}{\partial z}.
\end{equation}
Figure \ref{fig:BV_Lambda4} shows the height dependence of $\mathcal{N}^2$, scaled by the convective time scale $t_c = \sqrt{H/g \, \alpha \, \Delta T^2}$, for $Ra = 10^7$ and $\Lambda = 4$.
By definition in the stable region $\mathcal{N}^2 > 0$, and in the region where $0 \le \mathcal{N}^2  \le 1$ both entrainment and the internal gravity waves drive vertical motions of the fluid; internal gravity waves become dominant only for $z > 0.73$, where  $\mathcal{N}^2 > 1$.

\begin{figure}
\begin{centering}
\includegraphics[trim = 0 0 0 0, clip, width = \linewidth]{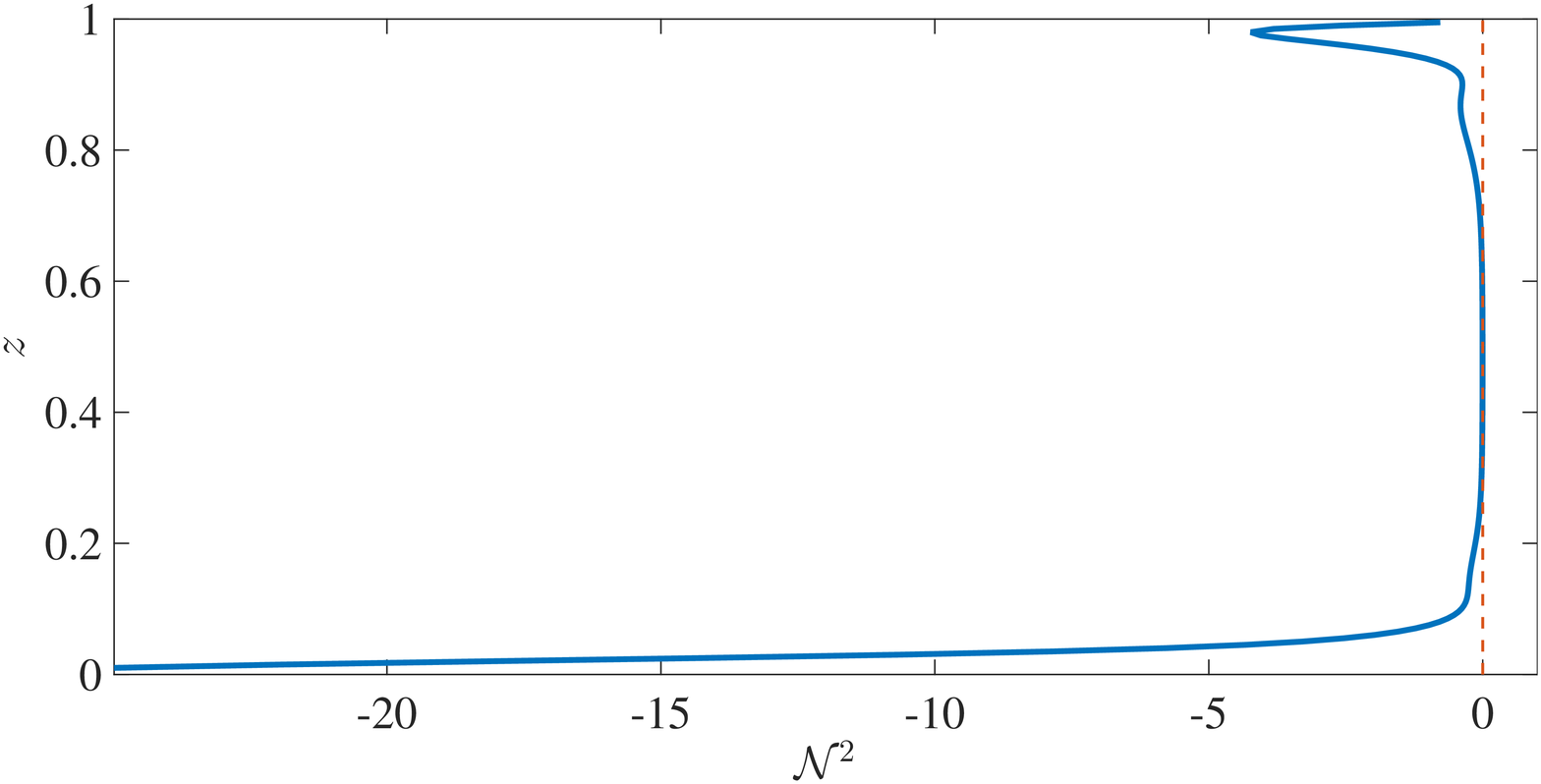} 
\caption{Variation of the Brunt-V\"{a}is\"{a}l\"{a} frequency, $\mathcal{N}^2$, with height for $Ra = 10^7$ and $\Lambda = 0.01$. The dashed vertical line is included to discern the change in sign in $\mathcal{N}^2$.}
\label{fig:BV_Lambda001}
\end{centering}
\end{figure}

In contrast, for $\Lambda = 0.01$, the internal gravity waves play no appreciable role in generating vertical motions. This can be seen from Figure \ref{fig:BV_Lambda001}, where $\mathcal{N}^2 < 0$ in the entire domain,  showing that convective motion of the fluid dominates.


\section{Conclusions}

We have systematically studied penetrative convection of a fluid with a density maximum using both analytical and numerical tools. We derived an evolution equation for the growth of the convecting layer by integrating the heat equation in the convecting layer and by constructing the mean temperature profile based on our knowledge of the flow in turbulent Rayleigh-B\'enard convection. In so doing, we have identified a new governing parameter, $\Lambda$, that measures the strength of the stratification of the stable upper layer and thereby exerts a controlling influence on the evolution of the underlying convecting layer. For a constant heat flux, $Q$, we recover the result from previous studies \cite{deardorff1969, tennekes1973, farmer1975} that the convecting layer grows diffusively.  The final steady thickness is shown to depend solely on the values of $Q$ and $\Lambda$.

In order to obtain an analytic equation for the evolution of $h(t)$, Eq. \ref{eqn:thickness_evolve}, we assumed that the heat transport in the stable layer is controlled by conduction. The veracity of this assumption is justified by the results of the analysis in the extreme limits of $\Lambda$, which provides the framework for the utility of such a simple approach.

High-resolution numerical simulations using the lattice Boltzmann method reveal that the growth of the convecting layer at a same $Ra$ depends sensitively on the value of $\Lambda$ -- the smaller the value of $\Lambda$, the faster the convecting layer grows. The flow field was also found to depend sensitively on $\Lambda$. For larger values of $\Lambda$, the penetrative entrainment of the plumes by the stable upper layer is suppressed. However, for smaller $\Lambda$ entrainment into the stable layer is efficient and the flow rapidly reaches a stationary state. The temporally and horizontally averaged temperature profile for $\Lambda = 4$ and $Ra = 10^7$ was found to be in qualitative agreement with the temperature profile from the experiments of Adrian \cite{adrian1975}.

We computed $Nu$ for $Ra = \left[10^6, 10^8\right]$ and $\Lambda = \left[0.01, 4\right]$ and found that for a fixed $Ra$, as $\Lambda$ decreases, $Nu$ increases. This is consistent with the limit of $\Lambda \rightarrow 0$ in penetrative convection reducing to that of the classical Rayleigh-B\'enard convection. For $\Lambda$ = $\left[0.01, 4\right]$, power-laws were obtained for the data using a linear least-squares fit, giving the exponent $\beta$ in $Nu = A \times Ra^{\beta}$. Both $A$ and $\beta$ vary non-monotonically with $\Lambda$, but a consistent physical interpretation is only possible by studying the changes in $Nu$ and not $A$ or $\beta$ individually.


We conclude by noting that whilst the complexities of many of the astrophysical and geophysical settings in which penetrative convection is operative are not at play in our study, nonetheless key qualitative phenomena will not differ.  Of principle relevance is the influence of rotation, which has the general effect of suppressing convection, as does stratification.  
Indeed, there is a direct mathematical analogy between rotating and stratified fluids, and under some conditions the analogy is exact \cite{Veronis:1970}.  Thus, because penetrative convection, as we have studied it here, couples a convective region with a strongly stratified region, we suggest that the analogy between rotation and stratification is of some use in considering the qualitative influence of rotation on our results.  

\begin{acknowledgements}
The authors acknowledge the support of the University of Oxford and Yale University, and the facilities and staff of the Yale University Faculty of Arts and Sciences High Performance Computing Center. 
S.T. acknowledges a NASA Graduate Research Fellowship.  J.S.W. acknowledges NASA Grant NNH13ZDA001N-CRYO, Swedish Research Council grant no. 638-2013-9243, and a Royal Society Wolfson Research Merit Award for support.
\end{acknowledgements}

\bibliographystyle{apsrev4-1}

%

\end{document}